\documentclass[12pt]{article}

\begin{document}

\title{Pulsars: Cosmic Permanent `Neutromagnets'?}
\author{Johan Hansson\footnote{c.johan.hansson@ltu.se} and Anna Ponga \\
 Department of Physics \\ Lule{\aa} University of Technology
 \\ SE-971 87 Lule\aa, Sweden}

\date{}

\maketitle

\begin{abstract}
We argue that pulsars may be spin-polarized neutron stars,
\textit{i.e.} cosmic permanent magnets. This would simply explain
several observational facts about pulsars, including the `beacon
effect' itself \textit{i.e.} the static/stable misalignment of
rotational and magnetic axes, the extreme temporal stability of
the pulses and the existence of an upper limit for the magnetic
field strength - coinciding with the one observed in ``magnetars".
Although our model admittedly is speculative, this latter fact
seems to us unlikely to be pure coincidence.
\end{abstract}

\section{Introduction}
We will assume that the simple model of a pulsar \cite{Bell} as a
rotating neutron star (NS) with a dipole magnetic field at an
angle with respect to its orbital axis \cite{Gold} is basically
correct. The radiated power from the magnetic dipole is
proportional to $sin^2 \theta$ \cite{Pacini}, where $\theta$ is
the angle between the dipole axis and and the rotational axis.

In order to make our point as simply as possible, we further
assume that:
\begin{itemize}
\item The NS is composed solely out of neutrons
\cite{BaadeZwicky}. (Nearly true assuming that quark stars do not
exist. There are observational indications \cite{NormalNS} that NS
indeed are composed out of normal nuclear matter.)

\item The density is constant throughout the NS and roughly the same
as the density of normal nuclear matter. (In reality, the density
is a few times higher in the NS core, and much less in its thin
crust.)

\item The magnetic field is due to spin-alignment of the neutrons
in the NS. This is motivated by the fact that aligned spins are
energetically favored by the nuclear force, as evidenced
\textit{e.g.} by the deuteron, the more strongly so for the
unusually small internucleon separation present in neutron stars
\cite{Nuclear}. We thus assume that the NS is a ``neutromagnetic"
material (in direct analogy to ferromagnetic materials). The
orbital angular momentum does not contribute to the magnetic field
as the neutrons are electrically neutral (no currents). We
understand that this is far from the orthodox view, however, the
extreme conditions inside neutron stars are not accessible to
direct experimental tests, so some leeway seems reasonable. The
Pauli principle, naively prohibiting parallel spin states for
$n-n$ (or $p-p$) may well be partially lifted by the extreme
gravitational and magnetic interactions, so that some quantum
numbers may differ\footnote{Note added in proof: Actually, the Pauli principle
is sidestepped as the neutrons occupy different positions in space, it is prohibitive
only for a fermion \textit{gas}, not a solid. The effect is well-known from terrestrial physics, for example there is
not much difference at low temperatures between a diamond made up of Carbon-12 (six protons, six neutrons and six electrons; bosons)
or Carbon-13 (six protons, seven neutrons and six electrons; fermions). The effect of the different statistics (Fermi or Bose)
is crucial only when particles can easily change places, in a solid (like our NS) it is in principle 
possible to ``tag'' a particle by its location; even though the spin-part of the wave function is identical for aligned spins
the spatial part is \textit{not}.}. Also, isotopic triplet ($I = 1$) states allow
spin triplet ($S = 1$) states for $n-n$ (and $p-p$). There are
also experimental observations of ferromagnet-like nuclear spin
ordering phenomena in controlled laboratory experiments
\cite{Koike} (first example of ``nuclear spin Ising system").

\end{itemize}

\section{Origin of magnetic field}
Magnetic fields generally can have two origins: i) charged
particles in motion, ii) alignment of magnetic moments of the
constituents.

The observationally inferred magnetic field of neutron stars range
from $10^4$ T for millisecond radio pulsars to a few times
$10^{11}$ T for magnetars.

There is no general consensus about the microscopic origin of the
magnetic field of a neutron star. If the ``lighthouse"/``beacon"
effect which produces the observed pulses in the assumed model
\cite{Gold} is correct, the magnetic field must be very strong and
at the same time very \textit{stable} to account for the fact that
pulsars are extremely accurate ``clocks". Any ``wobbling" or
dynamical behavior of the magnetic field would destroy the
accurate pulsing. The magnetic field must also be oriented in a
direction different from the rotational axis for any pulsar to
exist.

In our model, we automatically get all these characteristics, as
the neutron magnetic moments are ``frozen" in the same direction
by the requirement of lowest nuclear energy. In the orthodox
model, it is hard to see how a coupled [superfluid neutron -
superfluid and superconducting proton]-liquid can produce a
simple, and misaligned, dipole field, as a superconductor will
constrain the $B$-field into quantized vortex lines (and not give
rise to them). The electrons (expected to be ``normal") should be
electromagnetically coupled to the proton "fluid" and hence all
charged currents should co-rotate giving a magnetic field
collinear with the angular momentum. It is also known that several
dynamic magnetic instabilities may endanger the field itself. All
in all it seems that a more orthodox model of neutron star
interiors should give $B$-fields: i) collinear with $L$ ($\theta =
0$) and, ii) of highly dynamical complex non-dipole form.

In empirical nuclear potentials, \textit{e.g.} \cite{Nuclear}, it
can be seen that the spin-contribution becomes increasingly
attractive the smaller the separation. As the neutrons in a
neutron star are more highly packed than in normal nuclei, due to
gravitation, aligned spins are energetically favored
configurations.

Also, in the presence of gravity bound neutrons are stable. It
adds an additional, attractive background potential to the nuclear
one, lowering the potential below the level required for bound
states.

We take the attractive potential for aligned spins to be $\simeq
10$ percent of the total nuclear binding energy $\Delta m c^2$, as
corroborated by calculations in various models. (Roughly 0.1
$\times$ 10 MeV = 1 MeV or $10^{10}$ K.) The NS temperature,
originally also roughly $10^{10}$ K at birth in a supernova,
rapidly cools via the neutrinos produced in (gravity driven)
inverse beta-decay. When it falls below the neutron star
``Curie-temperature" $10^{10}$ K, the neutron star suddenly
becomes magnetized, the mechanism being analogous to the case in a
normal ferromagnetic material. If the temperature at creation
happens to be less than $10^{10}$ K the NS will be polarized from
the outset, the global energy minimum of the NS will correspond to
aligned neutron spins. In a NS the process is connected to the
strong nuclear force (instead of the electomagnetic force in a
ferromagnet). The NS can thus be labelled a ``neutro-magnetic"
material.

An independent way to motivate the numbers given above is to make
a calculation of the classical dipole-dipole interaction. Their
magnetic interaction energy is

\begin{equation}
E = \frac{\mu_0 \mu^2}{2 \pi x^3}
\end{equation}

where, for neutrons, $\mu = -1.91 \mu_N$ (the nuclear magneton),
$x \simeq 10^{-15}$ m (1 fm), giving $E \simeq 0.1$ MeV,
corresponding to a critical (``Curie") temperature of $T \simeq
10^9$ K. However, it is known that the above classical
dipole-dipole calculation underestimates the real value for iron
by almost four orders of magnitude, allowing for Curie
temperatures, and interaction energies, for `neutromagnets' to be
substantially higher. Also, quantum mechanical entanglement
effects should make the alignment much faster and more efficient,
due to quantum correlation occurring even at macroscopic
distances, as evidenced by laboratory experiments on the rate of
macroscopic magnetization due to entangled quantum state of
magnetic dipoles in salt \cite{Gosh}. As NS are expected to form
at $\sim 10^{10}$ K this could indicate that they become
magnetized already at birth, which may help explain the supernova
explosion itself, see below.

As all neutron stars seem to have very similar
masses\footnote{That this value coincides with the Chandrasekhar
limit, the maximum stable mass of a white dwarf, is a mystery in
itself.} $M_{NS} = 1.4 \pm 0.08 M_{\odot}$ \cite{NSmasses}, where
$ M_{\odot} = 1.99 \times 10^{30}$ kg is the solar mass (and from
general theoretical stability reasons cannot exceed $M_{NS} \sim 4
M_{\odot}$), we get for the maximum attainable permanent magnetic
field, corresponding to total, uniform polarization of the neutron
magnetic moments

\begin{equation}
B_{neutromagn} \leq 10^{12} \, T.
\end{equation}

This coincides nicely with the largest measured magnetic fields of
pulsars, in some so-called ``magnetars" \cite{Magnetars}. It seems
strange that such a close match should be pure coincidence.

\section{Origin of `beacon' effect}
The magnetic field of the massive progenitor star, especially in
its core, at the moment of collapse will tend to align the spins
of the nuclei, breaking spherical symmetry. As they come
sufficiently close the strong, spin dependent, nuclear force
suddenly becomes active, aligning the spins of the produced
neutrons in the same direction. The original magnetic field of the
star thus acts as a ``seed" for the final NS magnetic field (like
the magnetizing field in normal ferromagnetism). However, the
(``fossil") $B$-field of the original star is not conserved, and
boosted through contraction of the field lines, as most of the
star envelope is blown off. This is a problem in more orthodox
models especially in trying to reproduce the extreme $B$-fields of
magnetars \cite{Spruit}, but not in our case as it is known that
the magnetizing field can be a very small fraction (many orders of
magnitude) of the resulting permanent magnetic field. (The other
standard scenario, dynamo mechanism due to differential rotation
during collapse, seems destined to produce magnetic fields
collinear with the rotational axis, removing the `beacon´'
altogether.) We know from the sun that the magnetic field is not a
simple dipole, but has a more chaotic behavior (solar cycle, etc)
and does generally not coincide with the rotational axis. The
misalignment of the NS magnetic field will then be statistically
distributed with respect to its orbital axis, according to the
configuration at collapse. Also, the magnitude of the $B$-field
will be dependent on how complete the spin-polarization will be.
(Unless it always saturates, see section 6 below.) This, in turn,
will depend on the deviation from simple dipole at the time of
star collapse, differently polarized domains, etc.

In other models of neutron stars, where the interior is assumed to
consist of superfluid neutrons and superconducting protons
(roughly 1 percent of NS), it seems that the NS magnetic field
must lie along the orbital axis, which would preclude pulsars. The
superfluid neutron angular momentum vortices are strongly coupled
to the protons, creating strong magnetic fields parallel to the
orbital axis. If so, there would be no observable pulsars, as no
``beacon effect" results. In such models, the magnetic field is
believed to somehow arise in the highly (normal-) conducting
crust, but it is hard to see how it could reach the strength
\cite{Spruit}, stability and misalignment needed.

\section{Magnetic field - Period relation, and Glitches?}
Very fast, millisecond pulsars, generically seem to have the
weakest magnetic fields. In the orthodox view millisecond pulsars
are supposed to be old pulsars that have been spun-up by accretion
from a binary companion star. In our model one could imagine a
different scenario. The magnetic field is proportional to the
total spin of the neutrons, and only weakly dependent on other
variables,
\begin{equation}
B \propto S
\end{equation}
However, the orbital angular momentum is strongly dependent on
other variables, especially on the frequency of rotation, as the
mass and composition of the NS can be assumed to be fairly
generic,
\begin{equation}
L = L(\omega)
\end{equation}
The maximum angular momentum of a NS arising from
spin-polarization is
\begin{equation}
|\mathbf{S}| = N |\mathbf{s}_n| = N \frac{\hbar}{2} \simeq 10^{23}
Js,
\end{equation}
whereas the orbital angular momentum is a function of the
rotational angular frequency (or rotational period, $P$)
\begin{equation}
\mathbf{L} = I {\vec{\omega}} =  2 \pi \hat{\omega} I    / P
\end{equation}
The total angular momentum of the NS
is then
\begin{equation}
\mathbf{J} = \mathbf{L} + \mathbf{S}
\end{equation}
For a solitary (radio) pulsar, as there is no outside torque,
\begin{equation}
\frac{d \mathbf{J}}{dt} = 0.
\end{equation}

One could then speculate that pulsar glitches, sudden speed-ups of
$\Delta P/P \sim 10^{-8}$, may be due to rearrangement of
$\mathbf{L}$ and $\mathbf{S}$ through $L$-$S$ coupling, Tensor
coupling or relaxation (small amount of $\mathbf{S}
\leftrightarrow \mathbf{L}$). However, as pulsars exhibiting
glitches are very rare, the data set at present may be too small
to test such a hypothesis, and we will refrain from further
analysis here.

\section{Supernova explosion mechanism?}
\subsection{Supernova `bounce'}
In our model the NS (pulsar) could be the \textit{cause} of the
supernova (SN) explosion, and not an \textit{effect} of it. (At
the very least it will augment the explosion.) The `bounce' which
halts and reverses the infall of material may be due to
electromagnetic shock in the very dense plasma. This results when
the rapidly rotating NS and its dipole magnetic field is suddenly
born as a consequence of energy minimization. As $v_{sound}$ is
the speed of multi-nucleon interaction, and $v_{sound} \sim c$ in
the dense NS, this process takes only $R_{NS}/c \simeq 10^{-5}$ s,
quickly releasing the energy producing the huge magnetic field.

The electromagnetic forces can be seen to be more than enough for
the purpose of `bounce':
\begin{equation}
{\mathbf F} = q({\mathbf E} + {\mathbf v} \times {\mathbf B})
\end{equation}
For protons at the surface of a NS with $B \simeq 10^{12}$ T and
rotational period $P \sim 1$ s
\begin{equation}
 F_{EM}/F_{grav} \simeq 10^{13},
\end{equation}
and for electrons the ratio is $\sim 10^3$ higher. (Even if
combining the longest periods known, $P \sim 10$ s, with the
weakest inferred fields, $B \sim 10^4$ T, one obtains
$F_{EM}/F_{grav} \simeq 10^4$.)

In the conventional core-collapse scenario for a SN, the infall is
expected to ``bounce" (compression and rebound) when the inner
core exceeds nuclear densities and the, at \textit{very} short
distances, repulsive potential of the nuclear force ``stiffens"
the core. It is known that the outgoing shock-wave which results
is insufficient to disrupt the star. The shock stalls and the
material falls back onto the core. The usual way to remedy this is
by neutrino heating of the shock. However, even in this case it is
difficult to reproduce a star that actually explodes in
simulations \cite{CoreCollapse}. A pulsar-driven/augmented,
electromagnetic `bounce' would automatically give the
non-spherically symmetric explosion needed in core collapse SN
scenarios \cite{CoreCollapse}. Non-spherical SN ejecta are also
seen in observations \cite{SNObs}.

\subsection{Total explosion due to electromagnetic shock?}

Even though it is well known that 99 percent of the energy in a SN
is deposited in the neutrinos produced in inverse beta decay
driven by the gravitational potential during collapse, they may
have little to do with the explosion of the star, or at least not
be the dominating factor.

Energetically, the actual explosion of a SN is thus a minor
phenomenon ($\sim$ 1 percent of total).

$dB/dt$ is very large when neutron spins align. This, coupled with
the extremely dense plasma result in highly nonlinear plasma
interactions. The extremely strong, complexly entangled
electromagnetic field, give rise to turbulence and shocks. It also
rids angular momentum from the progenitor so that the central core
can rotate slower than break-up speed.

The theoretical break-up rotational period is $\sim 5 \times
10^{-4}$ s, but could be smaller during collapse when infalling
material stabilizes the core by external dynamical pressure, also
the proto-neutron star can then not be considered an isolated
object, invalidating the theoretical calculation of the limit
(gravitational radiation leaking angular momentum away).

The automatic deviation from spherical symmetry is also necessary
for explaining the observations of very high peculiar velocities
of many NS \cite{NSkicks}

Furthermore, the unexplained blow-off of the envelope of
non-massive stars \cite{White Dwarfs} can be due to the same
(universal) mechanism. It is less violent because the driving
force originates in a white dwarf with much lower $B$ and
$\omega$.

\subsection{Energy balance}
The energy released as the neutron spins align, assuming for now
100 percent (saturated) spin-polarization, is

\begin{equation}
\Delta E_{spin} \simeq \frac{N}{2} \times 1 \, MeV,
\end{equation}
where
\begin{equation}
N \simeq \frac{1.4 M_{\odot}}{m_n} \simeq 10^{57},
\end{equation}
the number of neutrons in a generic neutron star. (The actual
number is somewhat higher as the attractive potential lowers the
effective mass.)

So we get for the energy release due to spin-alignment
\begin{equation}
\Delta E_{spin} \simeq 10^{51} \, erg
\end{equation}
This energy is thus in principle capable of powering the whole SN
explosion. A canonical SN has a total energy output of roughly
$10^{51} \, erg$, but only 1 percent of this goes into the kinetic
energy of the ejecta, the rest (99 percent) escapes as neutrinos.

\section{Universal NS-`magnet'?}

Magnetic (dipole) field strengths of pulsars are
\textit{indirectly inferred} from observed spin-down rates,

\begin{equation}
B_{inferred} = (\frac{3 c^3 I}{8 \pi^2 R^6})^{1/2}(P
\dot{P})^{1/2},
\end{equation}
or, in Tesla,
\begin{equation}
B_{inferred} \simeq 10^{15}(P \dot{P})^{1/2},
\end{equation}
where $P$ is measured in seconds and $\dot{P} = dP/dt$ is
dimensionless.

In a normal ferro-magnet below the Curie-temperature the spin
alignment is near 100 percent. In a neutron star the process
should be at least equally efficient, and most likely also faster,
as it is driven by the strong nuclear force instead of
electromagnetism.

If we assume that the same (but with much higher effective binding
forces) applies for neutron stars, they will all be almost
identical permanent magnets. NS will then be extremely simple, all
having almost the same mass ($1.4 \pm 0.08 M_{\odot}$ from
observations \cite{NSmasses}) and the same magnetic field ($\sim
10^{12}$ T). This loss of individuality is well in line with the
next step on the cosmic compact object ladder, the black hole,
which is very simple and is totally described by only three
numbers (its mass $M$, angular momentum $L$ and charge $Q$).

If now $B$ is \textit{constant}, the power of dipole radiation
$dE/dt$ depends on angle and period only,
\begin{equation}
dE/dt = const \frac{sin^2 \theta}{P^4},
\end{equation}
where $const = 32 \pi^4 R^6 B^2/3 c^3$

In cases where $B$ is parallel to $L$ ($\theta = 0$) no pulsar
appears, if they are almost aligned ($\theta \sim 0$) a ``weak"
$B$ is inferred, and for large misalignment ($\theta \sim \pi /2$)
a huge ``magnetar" $B$ is inferred.

Assuming the pulsar-driven SN mechanism above, no SN should appear
in the $\theta = 0$ case. The massive star will then quietly
settle to a black hole as the energy dissipates, and gravity
overtakes everything else. The relative ``violence" and spatial
structure of a SN then depends only on the angle $\theta$, and
$P$.

\section{Conclusions}
Even though the presented model of a neutron star being a ``giant
polarized nucleus" is overly simplified, it nevertheless has an
attractive simplicity - in the vein of Zwicky, who together with
Baade originally introduced the very concepts of NS, SN and their
interconnections \cite{BaadeZwicky} - and explains several
unresolved properties of pulsars:

i) The origin of the magnetic field is simple and unavoidable. In
other models it is a complication which has to be addressed
separately. That a completely polarized neutron star automatically
gets a magnetic field comparable to that of magnetars seems, to
us, too compelling to be pure coincidence.

ii) The non-zero angle of the magnetic field to the rotational
axis is explained. The direction is triggered by the original
magnetic field of the massive star at time of collapse, and then
``frozen in" by the nuclear force.

iii) We get a natural maximum limit for the magnetic field, $B
\simeq 10^{12}$ T, corresponding to the field in `magnetars'. The
model also predicts that no pulsars (or neutron stars) will have a
$B$-field greater than this, as all measured neutron star masses
are highly peaked around 1.4 solar masses, and from general
stability arguments their maximum masses cannot be more than a few
times higher than this. In that sense our model is directly
falsifiable; if any neutron star with $B > 10^{12}$ T is detected
some other mechanism for generating the magnetic field must apply.

iv) The fact that pulsars are extremely exact ``clocks" means that
their magnetic fields must be very stable. As the neutrons align
their spin akin to the atoms in a normal ferromagnet, we get this
property for free.

v) Glitches may possibly be caused by relaxation, due to $L$-$S$
coupling, to a state with lower energy. This should then be
accompanied by a (minute) decrease in the $B$-field, which in
principle could be measured.

vi) If only the small proton admixture, of order 1 percent in the
orthodox scenario, contributes to permanent magnetization through
quantum mechanical $(n-p)$ pairing, $B_{max} \sim 10^{10}$ T, with
only marginal alteration in Curie temperature.

One should remember that the nuclear physics at these extreme
circumstances and densities is not known \textit{a priori}, so
several unexpected properties (such as ``neutromagnetism") might
apply. The ``proof is in the pudding", and from our
back-of-the-envelope calculations the model is at least not
immediately ruled out. The fact that there also exists a huge
``seeding" external magnetizing field from the collapsing star at
the moment of neutron star creation makes neutro-magnetization
plausible.

\end{document}